\documentstyle[preprint,epsfig,aps]{revtex} 
\draft
\begin{document}\tighten


\title{Evolution of cosmological  perturbations in a brane-universe}
\author{David Langlois}
\address{ Institut d'Astrophysique de Paris, \\
Centre National de la Recherche Scientifique,\\
98bis Boulevard Arago, 75014 Paris, France \\
and\\
D\'epartement d'Astrophysique Relativiste et de Cosmologie,\\
Centre National de la Recherche Scientifique,\\
Observatoire de Paris, 92195 Meudon Cedex, France }

\date{\today}

\maketitle

\def\beq{\begin{equation}}
\def\eeq{\end{equation}}
\def\d{{\delta}}
\def\4G{{\delta G^{st}}}
\def\5G{{\delta {}^{(5)}}G}

\begin{abstract}
The present article analyses the impact on cosmology, in particular on 
the evolution of cosmological perturbations, of the existence of 
extra-dimensions. The model considered here is that of a five-dimensional 
Anti-de Sitter spacetime where ordinary matter is confined to a brane-universe.
The homogeneous cosmology is recalled. The equations governing the evolution 
of cosmological perturbations are presented in the most transparent way: they 
are rewritten in a form very close to the equations of standard cosmology 
with two types of corrections: a. corrections due to the unconventional 
evolution of the homogeneous solution, which change the background-dependent 
coefficients of the equations; b. corrections due to the curvature along the 
fifth dimension, which act as source terms in the evolution equations.
\end{abstract}

\section{Introduction}

Although the idea of spatial extra-dimensions is an old one, it has recently
been  going through a revival  with the suggestion that ordinary matter is
 confined to a three-dimensional subspace, 
or {\it brane}, within a higher dimensional space or {\it bulk}. 
A consequence of this 
restriction, suggested by recent developments in string theory, is that usual 
Kaluza-Klein constraints on the size of the extra-dimensions are evaded, i.e. 
extra dimensions can be {\it large} or, in other words, the fundamental 
gravity mass scale, from which the usual Planck mass would 
be  derived, can be {\it 
low}. An interesting model,  due to Randall and Sundrum \cite{rs99b} and 
which has attracted a lot of attention, takes into 
account the warping effect on the five-dimensional geometry of the self
energy density of the brane and, provided one assumes the existence of a
negative cosmological constant in the bulk (which cancels the square of the 
brane energy density), one can find a static geometry with an infinite
 fifth dimension but   a finite Planck 
mass, in which usual gravity is 
recovered at first approximation \cite{gt}.

All these ideas have aroused a lot of interest in the particle physics 
community and search for extra-dimensions has been undertaken. 
However, if 
extra-dimensions are not so ``large", their  signature would be visible at 
energies that   are beyond collider reach (or on length scales too small 
to be probed by gravity experiments). This is why it might be important
to consider the effect of extra-dimensions in cosmology, where 
 very high energies have been reached in  the early universe and the 
modifications due to extra-dimensions could have left some imprint that 
might be observable, e.g. in the Cosmic Microwave Background (CMB) 
anisotropies.

Many works have been devoted to cosmology with extra-dimensions recently, 
essentially all in the context of {\it five-dimensional} spacetimes, i.e. 
with only one extra-dimension. 
In this framework,  it has  been realized \cite{bdl99}
 that one cannot recover the usual Friedmann equations 
in general, the case of an empty bulk, for instance,
 leading to a Hubble parameter which 
is proportional to the energy density of the brane. A way out has been 
found by applying the Randall-Sundrum idea to cosmology (\cite{cosmors}
 and \cite{bdel99}), i.e. considering
an Anti-de Sitter  bulk spacetime (with 
a negative cosmological constant $\Lambda$).  
One then has  to solve, 
assuming  homogeneity in the three ordinary spatial dimensions, 
 the  five-dimensional Einstein's equations $G_{AB}\equiv R_{AB}-R g_{AB}/2=
\kappa^2 T_{AB}$, where the energy-momentum tensor is composed of 
a cosmological constant contribution $-\Lambda g_{AB}$ and of an energy-momentum
tensor confined to the brane. This is equivalent to solve the Einstein's 
equations in the bulk, and then 
apply, at the brane location, junction conditions,
which relate the extrinsic curvature jump to the brane matter content.
Assuming in addition a planar symmetry, 
the  analog of  the Friedmann equation, 
relating the evolution of the brane-universe scale factor $a(t)$ 
 to its matter 
content,  has been found to be given by 
(\cite{kraus,bdel99})
\beq
H^2\equiv \left({\dot a\over a}\right)^2= 
{\kappa^4\over 36}\rho_b^2+{\Lambda\over 6},
\eeq
where $\rho_b$ is the total energy density of the brane (see also \cite{bcg00} for 
a more general approach). Decomposing it into 
a constant tension $\sigma$, such that $\kappa^4\sigma^2+6\Lambda=0$ 
(the Randall-Sundrum condition) and an `ordinary' cosmological energy density 
$\rho$, so that $\rho_b=\sigma +\rho$, one obtains 
\beq
H^2= {8\pi G\over 3}\left(\rho+{\rho^2\over 2\sigma}\right),
\label{new_friedmann}
\eeq
with the identification $\sigma=8\pi G$. This equation 
 gives the usual evolution in the
low energy regime $\rho\ll\sigma$ and quadratic corrections in the 
high energy regime $\rho \gtrsim \sigma$. In general, there is also an 
extra-term, which behaves effectively like a radiation-component and  
can be interpreted as a Schwarzschild-type mass (the bulk being then 
Schwarzschild-Anti de Sitter), but which will be neglected here.
 
With the above generalized law, all our understanding  of 
 {\it  homogeneous cosmology} is safe, provided  the universe has been in a 
low-energy regime 
since nucleosynthesis.  
A high-energy regime can be envisaged only 
in the earlier universe, for example during inflation \cite{mwbh99}.

The next step is obviously to investigate what will be the influence of 
extra-dimensions on the {\it cosmological perturbations} and their evolution. 
Several works have developed formalisms to handle the cosmological 
perturbations for a brane-universe in a five-dimensional spacetime 
\cite{mukohyama00,kis00,maartens00,l00,bdbl00,ks00,lmw00} but they 
are rather difficult to manipulate and their connection with the 
standard theory of cosmological perturbations remains somewhat obscure.
The aim of this  work is to present,  in the most transparent way,  
{\it how the  usual evolution equations for cosmological perturbations are 
modified in the braneworld scenario.}

\section{The modified perturbation equations}
 Let us start with a  perturbed five-dimensional 
metric, $g_{AB}=\bar g_{AB}+h_{AB}$, of the specific form 
\beq
ds^2=-n^2 (1+2A)dt^2+2n^2 \partial_i B dt dx^i 
+a^2\left[ (1+2C)\d_{ij}+2\partial_i \partial_j E\right]dx^i dx^j
+dy^2,
\eeq
where $t$ and the three $x^i$ correspond to the ordinary time and space 
dimensions, whereas $y$ denotes the fifth dimension;
$n(t,y)$ and $a(t,y)$ describe the background solution 
(see \cite{bdel99} for an explicit form), and $A$, $B$, $C$ and $E$ are four 
linear perturbations (that depend on all coordinates) of the scalar type 
(in the Bardeen \cite{bardeen} terminology). 
Note that restrictions on the coordinate systems have been assumed 
implicitly in order to get $h_{A5}=0$. Moreover, it will be assumed  
that  the (perturbed) brane is  at 
$y=0$, in other words we consider only Gaussian Normal (GN) coordinate systems.
This choice might not be the most convenient to solve for the bulk
 perturbations but it is very useful in order
 to make the link with  the familiar perturbations of standard cosmology.
Indeed the usual cosmological perturbations for the metric will be simply 
the values taken at $y=0$ of the five-dimensional fields $A$, $B$, $C$ and $E$.

Let us just recall that the equations for perturbations are obtained 
in standard cosmology by linearizing Einstein's  equations about a background
(homogeneous) solution. The linearized perturbation of the Einstein's tensor
are the basis of the standard theory of cosmological perturbations 
\cite{mfb,ks}:
\begin{eqnarray}
\4G_{00}&=& n^2\left(
6{\dot a\over a n^2} \dot C-{2\over a^2}\Delta C\right)
+[B,\dot E],\cr
\4G_{ij} &=& {a^2\over n^2}\left\{-2\ddot C+
\left(-6{\dot a\over a}+2{\dot n\over n}\right)\dot C 
+2{\dot a\over a}\dot A
+\Delta A+\Delta C 
+2\left({\dot a^2\over a^2}-2{\dot a \dot n\over an}+2{\ddot a\over a}\right)
\left(A-C\right)\right\}\d_{ij} \cr
&&  -\partial_i\partial_j\left(A+C\right)
+[B,\dot B, E, \dot E, \ddot E], \cr
\4G_{0i} &=&  \partial_i\left(-2 \dot C+2 {\dot a\over a}A\right)
+[B] ,  \label{einstein4}
\end{eqnarray}
where we have not written explicitly  the linear combinations
 of $B$ and $E$ and their 
time derivatives, which are summarized  by the brackets [in standard 
cosmology, it is possible to choose a coordinate system such 
that $B=E=0$].
The full evolution equations are then obtained by adding on the right hand side 
the energy-momentum tensor describing cosmological matter.

In brane cosmology, the way to obtain evolution equations similar to the 
standard equations is more tortuous and proceeds in several steps. 
The first step is to write the perturbed five-dimensional 
Einstein's tensor, which is 
more complicated than its four-dimensional counterpart in two respects, 
on one hand because there are five more components, on the 
other hand because the metric perturbations now depend on the fifth coordinate
$y$. Considering first the `ordinary' components, they can be written in the
form (see \cite{l00})
\begin{eqnarray}
\5G_{00}&=&  \4G_{00}+[h_\alpha, h'_\alpha,h''_\alpha]  \label{einstein5_00}\\
\5G_{ij} &=& \4G_{ij}+[h_\alpha, h'_\alpha,h''_\alpha]\label{einstein5_ij}\\
\5G_{0i} &=&  \4G_{0i}+[h_\alpha, h'_\alpha,h''_\alpha] , \label{einstein5_0i}
\end{eqnarray}
where the brackets stand for linear combinations of the four 
(scalar) metric perturbations
$h_\alpha=\{A,B,C,E\}$ ($\alpha=1,\dots, 4$), their derivatives with respect to $y$ and their 
second derivatives with respect to $y$. The coefficients in these linear 
combinations depend on the background solutions and involve time-derivatives
 as well as $y$-derivatives of the lapse and scale factor. The first term on 
the right hand side of each of the equations
 (\ref{einstein5_00}-\ref{einstein5_0i})
has exactly the same functional form as the corresponding 
right hand side in (\ref{einstein4})
but with the difference that the $h_\alpha$ are functions  not only of
the ordinary coordinates but of  $y$ as well [note that, now, we cannot
get rid of the terms involving $B$ or $E$ and their derivatives because we 
do not have the coordinate freedom to set these quantities to zero everywhere;
it is however possible to set $B=E=0$ on one slice, e.g. $y=0$, but, of course,
their y-derivatives cannot be eliminated]. 

Let us now evaluate the relations (\ref{einstein5_00}-\ref{einstein5_0i}) when $y\rightarrow 0$, i.e. 
when one goes {\it in the brane}. Second derivatives with respect to $y$ of $a$
and $n$  can be eliminated by resorting to the background 
Einstein's equations (which can be found in \cite{bdel99}). First derivatives 
of $a$ and $n$ can be replaced by the {\it background matter content} of the 
brane, via the junction conditions (see \cite{bdl99}):
\beq
\left({ a' \over a}\right)_{|y=0}=-\frac{\kappa^2}{6}\rho_b, 
\qquad
\left({ n' \over n}\right)_{|y=0}=\frac{\kappa^2}{6}  
\left( 3p_b + 2\rho_b \right), 
\label{junctions_b}
\eeq
where  $p_b=p-\sigma$ represents the total pressure of the brane, $p$ being the ordinary 
cosmological pressure.

Finally the first derivatives of the metric perturbations can be replaced 
by {\it the matter perturbations in the brane} according, once more, to the 
junction conditions (established in \cite{l00}):
\begin{eqnarray}
A'_{|y=0^+}&=& {\kappa^2\over 6}\left(2\d\rho+3\d p\right),\cr
B'_{|y=0^+}&=&\kappa^2(\rho+p){a_0\over n_0} v, \cr
\left(C'+{1\over 3}\Delta E'\right)_{|y=0^+}&=&-{\kappa^2\over 6}\d\rho, \cr
E'_{|y=0^+}&=& -2\kappa^2\pi^S,
\label{junctions_p}
\end{eqnarray}
where $v$ is the peculiar velocity potential and 
$\pi^S$ is related to the scalar part of the anisotropic stress tensor 
by $\pi_{ij}^S=(\partial_i\partial_j\pi^S-\Delta \pi^S \d_{ij}/3)$.

After substitution into (\ref{einstein5_00}-\ref{einstein5_0i}) of (\ref{junctions_b}) 
for the background coefficients and of (\ref{junctions_p}) for the perturbations, one
 arrives to the {\it modified equations governing the brane cosmological perturbations}, 
which are the  main result of this work:
\begin{eqnarray}
\4G_{00}-8\pi G\left[2\rho A+\d\rho\right]&=& 
8\pi G\left({\rho\over \sigma}\right)\left[\rho A+ \d\rho\right]
+3\theta_C+\Delta\theta_E, \label{evolution_00} \\
a^{-2}\4G_{ij}-8\pi G\left[\left(2p C+\d p\right)\d_{ij}+\pi^S_{ij}\right] &=&
 8\pi G\left({\rho\over \sigma}\right)\left[(2p+\rho)C\d_{ij}
+(\d p+(1+w)\d\rho)\d_{ij} \right.\cr
&& \left. -\left(1+3w\right)\pi^S_{ij}/2\right]
+\partial_i\partial_j\theta_E-\left(\Delta\theta_E+\theta_A+{2\over 3}\theta_C
\right)\d_{ij},  \label{evolution_ij} \\
\4G_{0i}+8\pi G\rho\left(1+w\right) a
\partial_i v &=&  4\pi G \left({\rho\over \sigma}\right)
\rho\left(1+w\right)\left(1+3w\right)a
\partial_i v+{1\over 2}\partial_i\theta_B, 
\label{evolution_0i}
\end{eqnarray}
in a coordinate system where $B(y=0)=E(y=0)=0$ (i.e. in the longitudinal 
gauge) and such that  $n(y=0)=1$. All five-dimensional quantities 
in (\ref{evolution_00}-\ref{evolution_0i}) are evaluated at $y=0$.
 We have defined $w=p/\rho$ and 
\beq
\theta_A={1\over n^2}\left(n^2 A'\right)', \quad
\theta_B={1\over n^2}\left(n^2 B'\right)', \quad
\theta_C={1\over a^2}\left(a^2 C'\right)', \quad
\theta_E={1\over a^2}\left(a^2 E'\right)'.
\eeq
In  (\ref{evolution_00}-\ref{evolution_0i}), we have isolated  
the corrective terms  on the right hand side, i.e. 
the standard equations correspond to the
 system (\ref{evolution_00}-\ref{evolution_0i})
 with zero on the right hand sides. 

It is then clear that there are {\it two types of corrective terms}. 
First, new terms arise because 
of the non-conventional nature of the generalized Friedmann law 
(\ref{new_friedmann}). One can see that 
these terms become negligible  in the low energy density regime, i.e. $\rho \ll
\sigma$. Second, there is another category of terms ($\theta_A$, 
$\theta_B$, $\theta_C$, $\theta_E$), which clearly come from the dependence 
of the metric perturbations on the fifth dimension. 
They represent the gradients of the 
metric perturbations along the fifth dimension, and, therefore, {\it all (scalar) 
 information about the outside of the brane is embodied in these four  four-dimensional fields}, 
which in fact are not independent as shown just below. 

At this stage, we have not yet taken into account the other components of the 
five-dimensional Einstein's equations, 
the $(5-5)$ component, and the (5-0) and (5-i) components. As far as the latter
are concerned, it is not very difficult to show that, via the same procedure
of substitution using (\ref{junctions_b}) and (\ref{junctions_p}), 
the component (5-0) is equivalent to the linearized energy conservation 
equation, whereas the components (5-i) yield the linearized cosmological 
Euler equation. These components therefore just correspond to  the usual 
conservation of the energy-momentum tensor. Finally, the (55) component, 
or rather its version in terms of the Ricci tensor, $\d R_{55}=0$, is
 expressible in the very simple form (see \cite{l00})
\beq
\theta_A+3\theta_C+\Delta \theta_E=0, \label{dR55}.
\eeq
 In view of (\ref{evolution_00}-\ref{evolution_0i}), it is then 
natural to reexpress these four quantities in terms of the 
following three  quantities,
\beq
8\pi G\,\rho_5= 3\theta_C+\Delta \theta_E=-\theta_A, 
\quad 
8\pi G\,\pi^S_5= \theta_E,
\quad 
8\pi G(\rho+p) a\, v_5= -\theta_B/2,
\eeq
which can be interpreted respectively as an effective energy density, 
an effective anisotropic stress and an effective peculiar velocity potential. 
The 
label `5' for these quantities refers to their `extra-dimensional' origin. 
One can also define an effective pressure, but 
which is not independent of $\rho_5$ because of (\ref{dR55}),
\beq
 P_5= -(8\pi G)^{-1} \left(\theta_A+2\theta_C+{2\over 3}\Delta\theta_E\right)
={1\over 3} \rho_5.
\eeq
It should not be surprising to find this equation of state, when one remembers
the existence of a radiation-like term in the homogeneous cosmology.
 All Einstein's equations have now been exhausted. Before  discussing 
influence of the corrective terms, let us just mention that 
the same procedure applies to vector and tensor perturbations. Moreover, 
it is straightforward to generalize the equations 
(\ref{evolution_00}-\ref{evolution_0i}) to a system of several fluids, 
by simply summing on all fluids in the right hand sides of (\ref{junctions_b})
and (\ref{junctions_p}).
These extensions as well as more details on the derivation of the evolution 
equations will be given elsewhere \cite{langlois}.

\section{Influence of the unconventional background corrections}

Let us  first concentrate on the  corrective terms that  just change 
the coefficients of the familiar evolution equations,  ignoring in this section
the corrections due to the curvature along the fifth dimension. If 
$\rho$ is negligible with respect to the tension $\sigma$, one recovers 
exactly the standard evolution for the perturbations. Deviations from the 
standard evolution will  appear in the high energy regime, which 
is possible only before nucleosynthesis. However, cosmological perturbations
of observational interest have entered the Hubble radius  long after nucleosynthesis,
which implies that possible modifications due to a high energy radiation era
have necessarily taken place while their lengthscale 
 was much bigger than the Hubble radius. 

A transition between a high energy regime and a low energy regime will 
thus affect all the relevant perturbations in the same way. Let us compute this 
`transfer coefficient'. 
 Assuming, for simplicity, that the anisotropic stress vanishes, 
the traceless part of 
(\ref{einstein5_ij}) yields $A=-C\equiv\Phi$, $\Phi$ being the usual 
Bardeen potential.  Combining then the  trace  of 
(\ref{einstein5_ij}) with (\ref{einstein5_00}), one 
obtains   the following evolution 
equation for the gravitational potential $\Phi$:
\beq
\ddot \Phi+\left[4+3c_s^2-6{\dot H\over \lambda^2\sigma}\right]H\dot \Phi
+\left[2\dot H+3H^2\left(1+c_s^2-2 {\dot H\over \lambda^2\sigma}\right)\right]
\Phi-\left(c_s^2-2 {\dot H\over \lambda^2\sigma}\right){\Delta \Phi\over a^2}=0,
\eeq
assuming a single barotropic fluid with sound speed $c_s^2=\dot p/\dot\rho$.
The only difference between this equation and 
 its analog in standard cosmology comes from the presence of terms
proportional to $\dot H/ \lambda^2\sigma$, terms which are negligible 
with respect to the others when one considers the low-energy regime. 

Taking the long wavelength limit of the above equation, and reexpressing
$c_s^2$ in terms of the scale factor only, one recovers, even in the 
high energy regime,  {\it exactly} the same 
equation as in standard cosmology  (which is not surprising if one adopts a 
purely geometric description of the perturbation), and its solution is given by 
\beq
\Phi=C_1\left(1-{H\over a}\int dt \, a\right)+C_2{H\over a},
\eeq
where $C_1$ and $C_2$ are two constants ($C_2$ corresponds to a decaying mode).
During the radiaton era, $a(t)\sim t^{1/4}$ in the high energy regime 
and $a(t)\sim t^{1/2}$ in the low energy regime so that the relation between 
the gravitational potentials in the two regimes is $\Phi$ (low energy)$=(5/6)
\Phi$(high energy). There is therefore a small attenuation of the 
perturbation amplitude during the high/low energy transition.

\section{Impact of the fifth dimension curvature corrections}
Let us turn now to the five dimensional curvature corrections.
It must be  clear that the 
three four-dimensional fields $\rho_5$, $\pi^S_5$ and $v_5$ are {\it arbitrary
fields from the four-dimensional point of view in the brane}, and that they 
can be determined only from a five-dimensional analysis of the bulk 
perturbations.
If it is not  difficult to compute  the perturbations in the 
Anti-de Sitter bulk, which are five-dimensional gravitational waves,  it is 
more delicate in general  to extract from all  possible bulk perturbations those
 which are  compatible with the brane and its perturbations. 

As far as we are interested here, perturbations from the bulk 
 appear in the four-dimensional-like 
equations for cosmological perturbations {\it like source terms}, and 
formally their impact on cosmological perturbations is similar to that 
of the `active seeds', which have been  studied  in the context of topological 
defects \cite{defects}. Part of the methods developed in order to investigate the effect 
of topological defects on CMB anisotropies could thus be converted to the 
study of brane-universe perturbations. In order to compute the CMB 
anisotropy power spectrum, one would need correlators between the various
quantities $\rho_5$, $v_5$, etc, which once more requires a specific description 
of the perturbations in the bulk, far beyond the scope of the present work. 

What can be infered, however,  from the various studies of topological 
defects, is that it seems unlikely that the `extra-dimension
seeds' may give the  dominant contribution in the CMB anisotropies. 
Indeed, active seeds were shown to fail to reproduce the gross 
features of the CMB anisotropies, in particular the acoustic peaks, because
they produce new metric perturbations at all times, thus blurring any 
specific features by cumulative effect. One can expect the same to happen 
with extra-dimension seeds. 
However,  they could constitute a subdominant contribution, that might be 
observable by high precision  CMB experiments in the near future. 
Therefore, an interesting question left for future 
investigation is wether it would be possible to {\it discriminate between 
topological defects and extra-dimension seeds}.

\end{document}